\documentclass[10pt,A4paper,conference]{IEEEtran}
\usepackage{amsmath}
\usepackage{amssymb}
\usepackage{upref}
\usepackage{amsfonts}
\usepackage{graphicx}
\usepackage{verbatim}

\newcommand{\field}[1]{\mathbb{#1}}
\newcommand{\C}{\field{C}}
\newcommand{\F}{\field{F}}

\newcommand{\cF}{{\cal F}}

\newcommand{\cG}{{\cal G}}

\newcommand{\cP}{{\cal P}}

\newcommand{\sP}{\cP}
\newcommand{\sG}{\cG}

\newcommand{\Gr}{\smash{{\sG\kern-1.5pt}_q\kern-0.5pt(n,k)}}
\newcommand{\Grtwo}{\smash{{\sG\kern-1.5pt}_2\kern-0.5pt(n,k)}}
\newcommand{\Gkone}{\smash{{\sG\kern-1.5pt}_q\kern-0.5pt(n,k_1)}}
\newcommand{\Gktwo}{\smash{{\sG\kern-1.5pt}_q\kern-0.5pt(n,k_2)}}
\newcommand{\Ps}{\smash{{\sP\kern-2.0pt}_q\kern-0.5pt(n)}}

\newtheorem{example}{Example}

\newtheorem{theorem}{Theorem}

\newtheorem{lemma}{Lemma}
\newtheorem{remark}{Remark}

\begin{document}

\title{Enumerative Encoding in the Grassmannian Space}

\author{\authorblockN{Natalia Silberstein}
\authorblockA{Department of Computer Science\\
Technion-Israel Institute of Technology\\
Haifa 32000, Israel \\
Email: natalys@cs.technion.ac.il} \and
\authorblockN{Tuvi Etzion}
\authorblockA{Department of Computer Science\\
Technion-Israel Institute of Technology\\
Haifa 32000, Israel \\
Email: etzion@cs.technion.ac.il}}

\maketitle
\begin{abstract}
Codes in the Grassmannian space have found recently application in
network coding. Representation of $k$-dimensional subspaces of
$\F_q^n$ has generally an essential role in solving coding
problems in the Grassmannian, and in particular in encoding
subspaces of the Grassmannian. Different representations of
subspaces in the Grassmannian are presented. We use two of these
representations for enumerative encoding of the Grassmannian. One
enumerative encoding is based on Ferrers diagrams representation
of subspaces; and another is based on identifying vector and
reduced row echelon form representation of subspaces. A third
method which combine the previous two is more efficient than the
other two enumerative encodings.
\end{abstract}

\section{Introduction}
\label{sec:introduction} Let $\F_q$ be a finite field of size $q$.
The\textit{ Grassmannian space} (Grassmannian, in short), denoted
by $\Gr$, is the set of all $k$-dimensional subspaces of  the
vector space~\smash{$\F_q^n$}, for any given two nonnegative
integers $k$ and $n$, $k \le n.$ A code $\C$ in the  Grassmannian
is a subset of $\Gr$.

Koetter and Kschischang~\cite{KK} showed the application of
error-correcting codes in  $\Gr$ to random network coding. This
application has motivated extensive work in the
area~\cite{XiFu07,EV08,MGR08,SKK08,GaYa08,EtSi08,Ska08}. On the
other hand, the Grassmannian and codes in the Grassmannian are
interesting for themselves~\cite{Knu71,Milne82,AAK,ScEt02,KoKu08}.
A natural question is how to encode/decode the subspaces in the
Grassmannian in  an efficient way. To answer this question we need
first to give a representation of subspaces, order all of them,
and encode/decode them based on this representation and order.

Cover~\cite{Cover} presented a general method of enumerative
encoding for a subset $S$ of binary words. Given a lexicographic
ordering of $S$, he presented an efficient algorithm for
calculating the index of any given element of $S$ (encoding). He
also presented  an inverse algorithm to find the element from $S$
given its index (decoding). Our goal in this paper is to apply
this scheme to all subspaces in a Grassmannian, based on different
lexicographic orders.

First, we present the  encoding  scheme of Cover~\cite{Cover}. Let
$\{0,1\}^n$ denote the set of all binary vectors of length $n$.
Let $S$ be a subset of $\{0,1\}^n$. Denote by
$n_S(x_1,x_2,\ldots,x_k)$ the number of elements of $S$ for which
the first $k$ coordinates are given by $(x_1,x_2,\ldots,x_k)$.

The lexicographic order is defined as follows. We say that for
$x,y\in \{0,1\}^n$, $x<y$, if $x_k<y_k$ for the least index $k$
such that $x_k\neq y_k$. For example, $00101<00110$.
\begin{theorem}\cite{Cover}
\label{thm:cover} The lexicographic index of $x\in S$ is
$$ind_S(x)=\sum_{j=1}^{n}x_j \cdot n_S(x_1,x_2,\ldots,x_{j-1},0).$$
\end{theorem}

\begin{remark}
The encoding algorithm of Cover is efficient if
$n_S(x_1,x_2,\ldots,x_{j-1},0)$ can be calculated efficiently.
\end{remark}

Let $S$ be a given subset and $i$ be a given index. The following
algorithm finds $x$ such that $ind_S(x)=i$.

\textbf{Inverse algorithm}~\cite{Cover}:
For $k=1,\ldots,n$, if $i\geq n_S(x_1,x_2,\ldots,x_{k-1},0)$ then
set $x_k=1$ and $i=i-n_S(x_1,x_2,\ldots,x_{k-1},0)$; otherwise set
$x_k=0$.

Cover~\cite{Cover} also presented the extension of these results
to arbitrary finite alphabet. For our purpose this extension is
more relevant as we will see in the sequel. The formula for
calculating the lexicographic index of $x\in
S\subseteq\{1,2,3,\ldots,M\}^n$ is  as follows.
\begin{equation}ind_S(x)=\sum_{j=1}^{n}\sum_{m<x_j}n_S(x_1,x_2,\ldots,x_{j-1},m).\label{cover}
\end{equation}
Cover didn't prove the correctness of this formula and didn't
present the inverse algorithm. We will present some of these
omissions for our decodings in the sequel.

In our work we present three different ways for enumerative
encoding of the Grassmannian. One is based on Ferrers diagrams
ordering; another is based on the identifying vectors combined
with the reduced row echelon form ordering; and the third one is a
combination of the first two.

The rest of this paper is organized as follows. In
Section~\ref{sec:form} we discuss different representations of
subspaces in the Grassmannian. We define the reduced row echelon
form of a $k$-dimensional subspace and its Ferrers diagram. These
two structures combined with the identifying vector of a subspace
will be our main tools for representation of subspaces. In
Section~\ref{sec:Ferrers} we define an order of the Grassmannian
based on Ferrers diagrams representation and present the first
enumerative encoding method. In Section~\ref{sec:Cover} we define
another lexicographic order of the Grassmannian based on
representation of a subspace by its identifying vector and its
reduced row echelon form and describe the second enumerative
encoding method. In Section~\ref{sec:combination}  we show how  we
can combine  two encoding methods mentioned above. Finally, in
Section~\ref{sec:conclude} we summarize our results and discuss
further applications of the different orders of the Grassmannian.
This leads for further results and problems for future research.

\section{Representation of Subspaces}
\label{sec:form}

In this section we give the definitions for two structures which
are useful in describing a subspace in $\Gr$, i.e., the reduced
row echelon form and the Ferrers diagram. The reduced row echelon
form is a standard way to describe a linear subspace. The Ferrers
diagram is a standard way to describe a partition of a given
positive integer. Based on these two structures and the
identifying vector of a subspace we will present a few
representations for subspaces which will be the key for our
enumerative encodings.

A $k$-dimensional subspace $X\in\mathbb F_q^n$ can be represented
by a $k\times n$ \textit{generator matrix} whose rows form a basis
for $X$. To have a unique representation of a subspace, we use the
following definition.

A $k \times n$ matrix with rank $k$ is in {\it reduced row echelon
form} (RREF in short) if the following conditions are satisfied.
\begin{itemize}
\item The leading coefficient of a row is always to the right of
the leading coefficient of the previous row.

\item All leading coefficients are {\it ones}.

\item Every leading coefficient is the only nonzero entry in its
column.
\end{itemize}

We represent a subspace $X$ of a Grassmannian by its generator
matrix in RREF. There is exactly one such matrix  and it will be
denoted by $RE(X)$.

\vspace{0.1cm}

\begin{example}
\label{exm:running} We consider the 3-dimensional subspace  $X$ of
$\F_2^7$ with the following eight elements.

\begin{footnotesize}
\begin{align*}
\begin{array}{cccccccc}
\text{1)} & (0 & 0 & 0 & 0 & 0 & 0 & 0) \\
\text{2)} & (1 & 0 & 1 & 1 & 0 & 0 & 0) \\
\text{3)} & (1 & 0 & 0 & 1 & 1 & 0 & 1) \\
\text{4)} & (1 & 0 & 1 & 0 & 0 & 1 & 1) \\
\text{5)} & (0 & 0 & 1 & 0 & 1 & 0 & 1) \\
\text{6)} & (0 & 0 & 0 & 1 & 0 & 1 & 1) \\
\text{7)} & (0 & 0 & 1 & 1 & 1 & 1 & 0) \\
\text{8)} & (1 & 0 & 0 & 0 & 1 & 1 & 0)
\end{array} .
\end{align*}
\end{footnotesize}

The generator matrix of $X$ in RREF is given by
\begin{footnotesize}
\begin{align*}
RE(X)=\left( \begin{array}{ccccccc}
1 & 0 & 0 & 0 & 1 & 1 & 0 \\
0 & 0 & 1 & 0 & 1 & 0 & 1 \\
0 & 0 & 0 & 1 & 0 & 1 & 1
\end{array}
\right) .
\end{align*}
\end{footnotesize}
\end{example}
\vspace{0.1cm}
\begin{remark}
It appears that designing an enumerative encoding of the
Grassmannian based on this representation won't be efficient and
we need to find other representations of a subspace for this
purpose.
\end{remark}

Each $k$-dimensional subspace $X$ of $\F_q^n$ has an {\it
identifying vector} $v(X)$~\cite{EtSi08}. $v(X)$ is a binary
vector of length $n$ and weight $k$, where the {\it ones} in
$v(X)$ are in the positions (columns) where $RE(X)$ has the
leading coefficients (of the rows).

\begin{remark}
We can consider an identifying vector $v(X)$ for some
$k$-dimensional subspace $X$ as a characteristic vector of a
$k$-subset. This coincides with the definition of rank- and
order-preserving map $\phi$ from $\Gr$ onto the lattice of subsets
of an $n$-set, given by Knuth~\cite{Knu71} and discussed by
Milne~\cite{Milne82}.
\end{remark}

\begin{example}
Consider the 3-dimensional subspace $X$ of
Example~\ref{exm:running}. Its identifying vector is
$v(X)=1011000$.
\end{example}
\begin{remark} For a representation of a $k$-dimensional
subspace $X$ we only need $v(X)$  and the $k\times (n-k)$ matrix
formed by the columns of $RE(X)$ which correspond to the
\textit{zeroes} in $v(X).$
\end{remark}

\begin{remark} A somewhat less compact way to represent a $k$-dimensional
subspace $X$ is to form a $(k+1)\times n$ matrix where the first
row is the identifying vector, $v(X),$  and the last $k$ rows form
the RREF of $X$, $RE(X)$. We will see in the sequel that this
representation will be very useful in our encoding algorithms.
\end{remark}
\vspace{0.01cm}

\begin{example}Consider the subspace $X$ of
Example~\ref{exm:running}. Its representation by a $(k+1)\times n$
matrix is given by

\begin{footnotesize}
\begin{align*}
\left( \begin{array}{ccccccc}
1 & 0 & 1 & 1 & 0 & 0 & 0\\
1 & 0 & 0 & 0 & 1 & 1 & 0 \\
0 & 0 & 1 & 0 & 1 & 0 & 1 \\
0 & 0 & 0 & 1 & 0 & 1 & 1
\end{array}
\right) .
\end{align*}
\end{footnotesize}
\end{example}
\vspace{0.3cm}

A \textit{partition} of a positive integer $m$ is a representation
of $m$ as  a sum of positive integers. The partition function
$p(m)$ is the number of partitions of $m$~\cite{vLWi92, AnEr04}.

\begin{example}
\label{exm:partition} One of the possible partitions of 21 is
$6+5+5+3+2$ and $p(21)=792.$
\end{example}
\vspace{0.15cm}

A {\it Ferrers diagram} $\cF$ represents a partition as a pattern
of dots with the $i$-th row having the same number of dots as the
$i$-th term in the partition~\cite{vLWi92,AnEr04}. A Ferrers
diagram satisfies the following conditions.
\begin{itemize}
\item The number of dots in a row is at most the number of dots in
the previous row.

\item All the dots are shifted to the right of the diagram.
\end{itemize}
Let $| \cF |$ denote the {\it size} of $\cF$, i.e., the number of
dots in $\cF$.

\vspace{0.01cm}
\begin{example}
\label{exm:diagram} For the partition of
Example~\ref{exm:partition} the Ferrers diagram $\cF$, $|\cF|=21$,
is given by

\begin{footnotesize}
\begin{align*}
\cF=\begin{array}{cccccc}
\bullet & \bullet & \bullet & \bullet & \bullet & \bullet \\
 & \bullet & \bullet & \bullet & \bullet & \bullet   \\
&\bullet & \bullet & \bullet & \bullet & \bullet  \\
&&&\bullet & \bullet & \bullet  \\
&&&&\bullet & \bullet
\end{array} .
\end{align*}
\end{footnotesize}
\end{example}
\vspace{0.3cm}

The {\it echelon Ferrers form} of a vector $v$ of length $n$ and
weight $k$, $EF(v)$, is the $k\times n$ matrix in RREF with
leading entries (of rows) in the columns indexed by the nonzero
entries of $v$ and $"\bullet"$  in all entries which do not have
terminal {\it zeroes} or {\it ones}. A $"\bullet"$ will be called
in the sequel a \textit{dot}. The dots of this matrix form the
Ferrers diagram of $EF(v)$. If we substitute elements of $\F_q$ in
the dots of $EF(v)$ we obtain a $k$-dimensional subspace $X$ of
$\Gr$. $EF(v)$ will be called also the echelon Ferrers form of
$X$. \vspace{0.01cm}

\begin{example}
\label{exm:echelonFerrersForm} The echelon Ferrers form of the
vector $v=1011000$ is

\begin{footnotesize}
\begin{align*}
EF(v)=\left( \begin{array}{ccccccc}
1 & \bullet & 0 & 0 & \bullet & \bullet & \bullet \\
0 & 0 & 1 & 0 & \bullet & \bullet & \bullet \\
0 & 0 & 0 & 1 & \bullet & \bullet & \bullet
\end{array}
\right)~.
\end{align*}
\end{footnotesize}\\
\end{example}
The {\it Ferrers tableaux form} of a subspace $X$, denoted by
$\cF(X)$, is obtained by assigning the values of $RE(X)$ in the
Ferrers diagram of $EF(v(X))$.
\begin{remark}
$\cF(X)$ defines a representation of $X$.
\end{remark}

\begin{example}
\label{exm:FerrersForm} For the subspace $X$, given in
Example~\ref{exm:running} whose echelon Ferrers form given
in~\ref{exm:echelonFerrersForm}, the Ferrers tableaux form is

\begin{footnotesize}
\begin{align*}
\cF(X)=
\begin{array}{cccc}
0 & 1 & 1 & 0 \\
&1 & 0 & 1  \\
&0 & 1 & 1
\end{array}.
\end{align*}
\end{footnotesize}\\
\end{example}
\vspace{-0.8cm}
\section{Encoding based on Ferrers Tableaux Forms}
\label{sec:Ferrers} In this section we present an encoding of the
Grassmannian based on the Ferrers tableaux form representation of
$k$-dimensional subspaces. The number of dots in a Ferrers diagram
of a $k$-dimensional subspace is at most $k \cdot (n-k)$. It can
be embedded in a $k\times (n-k)$ box. We define a lexicographic
order of such Ferrers diagrams, which induces an order of
subspaces in the Grassmannian, and then apply the enumerative
encoding to all $k$-dimensional subspaces.

The order that we  define in the sequel is based on the following
theorem~\cite{vLWi92} which shows the connection between the
number of $k$-dimensional subspaces of $\F_q^n$, denoted by the
$q$-ary Gaussian coefficient $\left[\begin{array}{c}
n\\
k\end{array}\right]_{q}$, and partitions.
\begin{theorem}
\label{thm:vanLint} For any given integers $k$ and $n$, $k \leq
n$,
\[ \left[\begin{array}{c}
n\\
k\end{array}\right]_{q}=\sum_{\ell=0}^{k(n-k)}\alpha_{\ell}q^{\ell},\]
where the coefficient $\alpha_{\ell}$ is the number of partitions
of $\ell$ whose Ferrers diagrams fit in a box of size $k \times
(n-k)$.
\end{theorem}
\subsection{Encoding of Ferrers Diagrams}
Let $\cF$ be a Ferrers diagram  of size $m$ embedded in  a
$k\times (n-k)$ box. We represent $\cF$ by an integer vector of
length $n-k$, $(\cF_{n-k},...,\cF_2,\cF_1),$ where $\cF_i$ is
equal to the number of dots in the $i$-th column of $\cF,$ $1\leq
i\leq n-k$, where we number the columns from right to left. Note
that $\cF_{i+1} \leq \cF_i$, $1 \leq i \leq n-k-1$.

Let $\cF$ and $\widetilde{\cF}$ be two Ferrers diagrams of the
same size. We say that $\cF < \widetilde{\cF}$ if $\cF_i
>\widetilde{\cF}_i$ for the least index $i$ such that  $\cF_i \neq
\widetilde{\cF}_i$, i.e., in the least column where they have a
different number of dots, $\cF$ has more dots than
$\widetilde{\cF}$.

Let $N_m(\cF_j,...,\cF_2,\cF_1)$ be the number of Ferrers diagrams
of size $m$ embedded in a $k\times (n-k)$ box, for which the first
$j$ columns are given by $(\cF_j,...,\cF_2,\cF_1)$. The number of
dots in column $j$ of $\cF$ is at most $\cF_{j-1}$. Hence, by
(\ref{cover}) the lexicographic index $ind_m$ of $\cF$ among all
the Ferrers diagrams with the same size $m$ is given by
\begin{equation}
ind_m(\cF)=\sum_{j=1}^{n-k}
\sum_{a=\cF_j+1}^{\cF_{j-1}}N_m(a,\cF_{j-1},...,\cF_2,\cF_1),\label{eq:ind_m}
\end{equation}
where we define $\cF_{0}=k.$

Note that $0\leq ind_m(\cF) \leq \alpha_{m}-1$, where $\alpha_{m}$
is defined in Theorem~\ref{thm:vanLint}.

Let $p(m,k,\eta)$ be the number of Ferrers diagrams of size $m$
which are embedded in a $k\times \eta$ box, i.e.,
$p(m,k,n-k)=\alpha_m$. The following lemma can be easily verified.
\begin{lemma}
\label{lem: recursion} $p(m,k,\eta)$ satisfies the following
recurrence relation:
$$p(m,k,\eta)=p(m-k, k, \eta-1)+p(m,k-1,\eta)$$
$$p(m,k,\eta)=p(k\eta-m,k,\eta),$$
with the initial conditions
\begin{eqnarray*}
&p(m,k,\eta)=0, \textrm{if }  m<0;\;\\
&p(m,1,\eta)=1, \textrm{if  }  m \leq \eta;\\
&p(m,k,1)=1, \textrm{if  }  m \leq k.
\end{eqnarray*}
\end{lemma}
\vspace{0.2cm}

\begin{remark}
Since $p(m,k,\eta)=p(k\eta-m,k,\eta),$ we can assume that $m\leq
\frac{k \eta }{2}$.
\end{remark}
Now, using the  definition of $p(m,k,\eta)$ we can calculate the
size of $N_m(\cF_j,...,\cF_2,\cF_1)$.

\begin{lemma}
\label{lem:calc_Nm}
\begin{equation*} N_m(\cF_j,...,\cF_2,\cF_1)=p(m-\sum_{i=1}^j\cF_i, \cF_j,n-k-j).
\end{equation*}
\end{lemma}

Lemma~\ref{lem:calc_Nm} implies that if we can calculate
$p(m,k,\eta)$ efficiently then we can calculate efficiently
$ind_m(\cF)$ for Ferrers diagram of size $m$ embedded in a
$k\times (n-k)$ box.

Given an index $i$, in a similar way to the inverse algorithm of
Cover we can design an inverse algorithm to find the Ferrers
diagram $\cF$ such that $i=ind_m(\cF)$.

Now, we can define an order of all Ferrers diagrams embedded in a
$k \times (n-k)$ box.

For two Ferrers diagrams $\cF$ and $\widetilde{\cF}$, we say that
$\cF < \widetilde{\cF}$ if one of the following conditions holds
\begin{itemize}
\item $|\cF| > |\widetilde{\cF}|$

\item $|\cF| = |\widetilde{\cF}|$,  and  $ind_{|\cF|}(\cF)<
ind_{|\widetilde{\cF}|}(\widetilde{\cF})$. \end{itemize}

\begin{example}
\label{ex:order_F} For the three Ferrers diagrams $\cF$,
$\widetilde{\cF}$, and $\widehat{\cF}$
\begin{footnotesize}
\begin{align*}
\cF = \begin{array}{ccc}
\bullet & \bullet & \bullet  \\
\bullet & \bullet & \bullet  \\
&&\bullet
\end{array},
 \: \widetilde{\cF} =\begin{array}{ccc}
\bullet & \bullet & \bullet  \\
&\bullet & \bullet   \\
&\bullet & \bullet
\end{array},
 \:\widehat{\cF} =\begin{array}{ccc}
\bullet & \bullet & \bullet  \\
&\bullet & \bullet   \\
&&\bullet
\end{array},
\end{align*}
\end{footnotesize}\\
we have $ \widetilde{\cF} < \cF < \widehat{\cF} .$
\end{example}

\subsection{Order based on the Ferrers Tableaux Forms}

Let $X,\: Y\in\Gr$ be two $k$-dimensional subspaces, $RE(X)$ and
$RE(Y)$ the related RREFs. Let $v(X)$ and $v(Y)$ be the
identifying vectors of $X$ and $Y$, respectively, and $\cF_X$,
$\cF_Y$ the related Ferrers diagrams of $EF(v(X))$ and $EF(v(Y))$.
Let $x_1,x_2,...,x_{|\cF_X|}$ and $y_1,y_2,...,y_{|\cF_Y|}$ be the
entries of Ferrers tableaux forms $\cF(X)$ and $\cF(Y),$
respectively. The entries of a Ferrers tableaux form are numbered
from right to left, and from top to bottom.

We say that $X < Y$ if one of the following conditions holds
\begin{itemize}
\item $\cF_X < \cF_Y; $

\item $\cF_X = \cF_Y $,  and $(x_1,x_2,...,x_{|\cF_X|}) <
(y_1,y_2,...,y_{|\cF_Y|}).$
\end{itemize}

\begin{example}
Let $X,Y,Z,W\in\mathcal G_2(6,3)$ which are given by

\begin{footnotesize}
\begin{align*}
RE(X)=\left(\begin{array}{cccccc}
1 & 0 & \textbf{1} & \textbf{1} & 0 & \textbf{1} \\
0 & 1 & \textbf{1} & \textbf{1} & 0 & \textbf{1} \\
0 & 0 & 0 & 0 & 1 & \textbf{1} \\
\end{array}
\right),~~ \cF(X)=
\begin{array}{ccc}
1 & 1 & 1 \\
1 & 1 & 1  \\
&  & 1
\end{array},
\end{align*}
\end{footnotesize}
\begin{footnotesize}
\begin{align*}RE(Y)=\left( \begin{array}{cccccc}
1 & \textbf{1} & 0 & 0 & \textbf{0} & \textbf{1} \\
0 & 0 & 1 & 0 & \textbf{0} & \textbf{0} \\
0 & 0 & 0 & 1 & \textbf{1} & \textbf{1}
\end{array}
\right) ,~~ \cF(Y)=
\begin{array}{ccc}
1 & 0 & 1 \\
& 0 & 0  \\
& 1 & 1
\end{array},
\end{align*}
\end{footnotesize}
\begin{footnotesize}
\begin{align*}
RE(Z)=\left( \begin{array}{cccccc}
1 & \textbf{1} & 0 & \textbf{1} & 0 & \textbf{1} \\
0 & 0 & 1 & \textbf{1} & 0 & \textbf{1}\\
0 & 0 & 0 & 0 & 1 & \textbf{0}
\end{array}
\right),~~ \cF(Z)=
\begin{array}{ccc}
1 & 1 & 1 \\
& 1 & 1  \\
& & 0
\end{array},
\end{align*}
\end{footnotesize}
\begin{footnotesize}
\begin{align*}
RE(W)=\left( \begin{array}{cccccc}
1 & \textbf{1} & 0 & \textbf{1} & 0 & \textbf{1} \\
0 & 0 & 1 & \textbf{1} & 0 & \textbf{1} \\
0 & 0 & 0 & 0 & 1 & \textbf{1}
\end{array}
\right),~~ \cF(W)=
\begin{array}{ccc}
1 & 1 & 1 \\
& 1 & 1  \\
& & 1
\end{array}.
\end{align*}
\end{footnotesize}
From Example~\ref{ex:order_F} we have $\cF_Y < \cF_X < \cF_Z
=\cF_W $. Since $(z_1,z_2,...,z_{|\cF_Z|})=(1,1,0,1,1,1)<
(w_1,w_2,...,w_{|\cF_W|})=(1,1,1,1,1,1)$ it follows that $Y < X <
Z < W$.
\end{example}

\vspace{-0.25cm}
\subsection{Encoding Based on the Ferrers Tableaux Forms}
Now, we use the order defined above and Theorem~\ref{thm:vanLint}
for enumerative encoding of $\Gr$. Let $\{x\}$ be the integer
value of vector $x=(x_1,...,x_{|\cF_X|})$ and let $\{i\}_{q}$ be
the base $q$ representation of the integer $i$.
\begin{theorem}
\label{thm:Ferrers_index} Let $X \in \Gr$, $\cF_X$ be the Ferrers
diagram of $EF(v(X))$,   $x_1,x_2,...,x_{|\cF_X|}$ be the entries
of $\cF(X)$. Then the index $Ind_1 (X)$, by the order based on the
Ferrers tableaux forms, is given by
\begin{equation*}Ind_1(X)=\sum_{i=|\cF_X|+1}^{k(n-k)}\alpha_{i}q^{i}+ ind_{|\cF_X|}(\cF_X)q^{|\cF_X|}+\{x\},\label{Ind1}
\end{equation*}
where $\alpha _i$ is defined in  Theorem~\ref{thm:vanLint} and
$ind_{|\cF_X|}$ is given by (\ref{eq:ind_m}).
\end{theorem}
\vspace{0.1cm}

Now, an index $i$ is given. The following algorithm returns a
subspace $X \in \Gr$ such that $Ind_1(X)=i$.

\textbf{Inverse algorithm:}

\textit{Step 1:} If $i< q^{k(n-k)}$ then $|\cF_X|=k(n-k)$; assign
the values of $\{i\}_q$ to $\cF(X)$ and stop; otherwise set
$i=i-q^{k(n-k)}$.

\textit{Step 2:} For $1\leq j\leq k(n-k),$ if $i<
\alpha_{k(n-k)-j}q^{k(n-k)-j},$ then  $|\cF_X|=k(n-k)-j$,
$\cF_X=ind_{|\cF_X|}^{-1}(\lfloor\frac{i}{q^{k(n-k)-j}}\rfloor)$;
assign the values of $\{i-\lfloor\frac{i}{q^{k(n-k)-j}}\rfloor
q^{k(n-k)-j}\}_q$ to $\cF(X)$ and stop; otherwise set
$i=i-\alpha_{k(n-k)-j}q^{k(n-k)-j}.$

\begin{theorem}
The complexity of the encoding/decoding based on the Ferrers
tableaux forms is $O(k^{5/2}(n-k)^{5/2}).$
\end{theorem}
\section{RREF and Identifying Vector Encoding }
\label{sec:Cover} In this section we provide another method for
enumerative encoding of the Grassmannian, based on the
representation of a subspace $X \in \Gr$ by a $(k+1) \times n$
matrix whose first row is $v(X)$ and the other $k$ rows form
$RE(X)$. First, we define the lexicographic order in the
Grassmannian based on this representation and then we apply
enumerative encoding to the Grassmannian based on this
representation.

\vspace{-0.1cm}

\subsection{Order based on the Extended Representation }
\label{ssec:lex_order}

Let $X\in\Gr$ be a $k$-dimensional subspace. The \textit{extended
representation} $EXT(X)$ of $X$ is a  $(k+1)\times n$ matrix
obtained by combining the identifying vector
$v(X)=(v(X)_n,\ldots,v(X)_1)$ and the RREF
$RE(X)=(X_n,\ldots,X_1)$, as follows
\begin{align*}
EXT(X)=\left( \begin{array}{cccc}
 v(X)_n & \ldots & v(X)_2 & v(X)_1 \\
 X_n & \ldots & X_2 & X_1
\end{array}
\right).
\end{align*}
Note, that $v(X)_i$ is the most significant bit of the column
vector $\begin{footnotesize} \left( \begin{array}{c} v(X)_i \\X_i
\end{array}\right) \end{footnotesize}$.

Let $X,Y\in \Gr$ and $EXT(X)$, $EXT(Y)$ be the extended
representations of $X$ and $Y$, respectively. Let $i$ be the least
index such that $EXT(X)$ and $EXT(Y)$ have different columns. We
say that $X<Y$ if $\begin{footnotesize}\left\{ \begin{array}{c}
v(X)_i \\X_i
\end{array}\right\}<\left\{ \begin{array}{c} v(Y)_i \\Y_i
\end{array}\right\}.\end{footnotesize}$

\begin{example}
For $X,Y\in\mathcal G_2(6,3)$ whose $EXT(X),$ and $EXT(Y)$ are
given by

\begin{footnotesize}
\begin{align*}
EXT(X)=\left(\begin{array}{cccccc}
1 & 1 & 1 & 0 & 0 & 0 \\
1 & 0 & 0 & 0 & 1 & 0 \\
0 & 1 & 0 & 0 & 0 & 0 \\
0 & 0 & 1 & 1 & 0 & 0 \\
\end{array}
\right)\end{align*}
\end{footnotesize}
\begin{footnotesize}
\begin{align*}
EXT(Y)=\left( \begin{array}{cccccc}
1 & 1 & 0 & 0 & 1 & 0 \\
1 & 0 & 0 & 0 & 0 & 0 \\
0 & 1 & 0 & 0 & 0 & 0 \\
0 & 0 & 0 & 0 & 1 & 0
\end{array}
\right) ,
\end{align*}
\end{footnotesize}
we have $X<Y$.
\end{example}

\subsection{Enumerative Encoding Based on Extended Representation}
\label{ssec:enumaration}

Let $\begin{footnotesize} N\left( \begin{array}{ccc}
 v_j & \ldots &  v_1 \\
 X_j & \ldots &  X_1
\end{array}
\right)
\end{footnotesize}$
be the number of elements in $\Gr$  for which the  first $j$
columns in the extended representation are given by
$\begin{footnotesize} \left( \begin{array}{ccc}
 v_j & \ldots &  v_1 \\
 X_j & \ldots &  X_1
\end{array}
\right).
\end{footnotesize}$

\begin{remark} We view all the $q$-ary vectors of length $k+1$
as our finite alphabet. Let $S$ be the set of all $q$-ary
$(k+1)\times n$ matrices which form extended representations of
some $k$-dimensional subspaces. Now, we can use Cover's method to
encode the Grassmannian. In this setting note that
$\begin{footnotesize} N\left( \begin{array}{ccc}
 v_j & \ldots &  v_1 \\
 X_j & \ldots &  X_1
\end{array}
\right)
\end{footnotesize}$ is equivalent to
$n_S(x_1,x_2,\ldots,x_j)$, where $x_i = \left( \begin{array}{c}
 v_i \\
 X_i
\end{array}
\right)$.
\end{remark}

\begin{lemma}
\label{lem:N}
\[
N\left( \begin{array}{ccc}
 v_j & \ldots &  v_1 \\
 X_j & \ldots &  X_1
\end{array}\right)=\left[\begin{array}{c}
n-j\\k-\sum_{i=1}^{j}v_{i}\end{array}\right]_{q}.
\]

\end{lemma}

\begin{theorem}
\label{thm:Ind} Let $X\in\mathcal{G}_{q}(n,k)$ be represented by
\[
EXT(X)=\left( \begin{array}{cccc}
 v_n & \ldots & v_2 & v_1 \\
 X_n & \ldots & X_2 & X_1
\end{array}
\right).
\]
Then the lexicographic index of $X$ is given by
\begin{equation*}
Ind_2(X)  =
\sum_{j=1}^{n}(v_{j}q^{k-w_{j-1}}+(1-v_{j})\frac{\left\{
X_{j}\right\} }{q^{w_{j-1}}})\left[\begin{array}{c}
n-j\\
k-w_{j-1}\end{array}\right]_{q},\label{Ind2}
\end{equation*}
where $w_{j-1}$ denotes the weight of the first rightmost $j-1$
entries of $v(X)$, i.e., $w_{j-1}=\sum_{\ell=1}^{j-1}v_\ell$.
\end{theorem}
\vspace{0.1cm}
\begin{example}
\label{exm:X_0} Let $X\in \mathcal G_2(6,3)$ be given by
\begin{footnotesize}
\begin{align*}
EXT(X)=\left( \begin{array}{cccccc}
0 & 1 & 0 & 1 & 1 & 0 \\
0 & 1 & 1 & 0 & 0 & 1 \\
0 & 0 & 0 & 1 & 0 & 0 \\
0 & 0 & 0 & 0 & 1 & 1
\end{array}
\right) .
\end{align*}
\end{footnotesize}
By Theorem~\ref{thm:Ind} we have
$$Ind_2(X)=5\cdot\left[\begin{array}{c}5\\3\end{array}\right]_{2}
+2^3\cdot\left[\begin{array}{c}4\\3\end{array}\right]_{2}
+2^2\cdot\left[\begin{array}{c}3\\2\end{array}\right]_{2}$$
$$+1\cdot\left[\begin{array}{c}2\\1\end{array}\right]_{2}
+2\cdot\left[\begin{array}{c}1\\1\end{array}\right]_{2}=928.
$$
\end{example}
\vspace{0.3cm}

Now suppose that an index $i$ is given. The following algorithm
finds a subspace $X \in \Gr$ such that $Ind_2(X)=i$.
\vspace{0.1cm}

\textbf{Inverse algorithm:} Set $i_{0}=i$.
\vspace{0.07cm}

\noindent For $j=1,2,...,n$ do:

\begin{itemize}
\item if $w_{j-1}\geq k$ then set $v(X)_{j}=0$, $X_{j}=\{0\}_q$,
and $i_j=i_{j-1}$;

\item otherwise
\begin{itemize}
\item if $i_{j-1}\geq q^{k-w_{j-1}}
\begin{footnotesize}\left[\begin{array}{c}
n-j\\k-w_{j-1}\end{array}\right]_{q}\end{footnotesize}$ then set
$v(X)_{j}=1$, $X_{j}=\{q^{w_{j-1}}\}_{q}$, and
$i_{j}=i_{j-1}-q^{k-w_{j-1}}\begin{footnotesize}
\left[\begin{array}{c}
n-j\\k-w_{j-1}\end{array}\right]_{q}\end{footnotesize}$; \item
otherwise let $val=\left\lfloor
i_{j-1}/\begin{footnotesize}\left[\begin{array}{c}
n-j\\k-w_{j-1}\end{array}\right]_{q}\end{footnotesize}\right\rfloor
$ and set $v(X)_{j}=0$, $X_{j}=\left\{ val \cdot
q^{w_{j-1}}\right\} _{q}$, and $i_{j}=i_{j-1}-val \cdot
\begin{footnotesize}\left[\begin{array}{c}
n-j\\k-w_{j-1}\end{array}\right]_{q}\end{footnotesize}.$
\end{itemize}
\end{itemize}

\begin{theorem}
The complexity of the encoding/decoding based on the extended
representation is $O(nk(n-k)\log n\log\log n).$
\end{theorem}

\section{Combination of the Encoding Methods}
\label{sec:combination}

The only disadvantage of the Ferrers tableaux form encoding is the
computation of the $\alpha_i$'s and $ind_{|\cF_X|}(\cF_X)$ in
Theorem~\ref{thm:Ferrers_index}. This is the reason for its
relatively higher complexity. The advantage of this encoding is
that once these values are known, the algorithm becomes trivial.
Our solutions for the computation of the $\alpha_i$'s and
$ind_{|\cF_X|}(\cF_X)$ are relatively not efficient and this is
the main reason why we turned to enumerative encoding based of the
RREF and the identifying vector of a subspace. The only
disadvantage of this enumerative encoding is the computation of
the Gaussian coefficients in Theorem~\ref{thm:Ind}. It appears
that a combination of the two methods is more efficient from the
efficiency of each one separately. The complexity will remain
$O(nk(n-k)\log n\log\log n)$, but the constant will be
considerably reduced in the average. This can be done if there
won't be any need for the computation of the $\alpha_i$'s and the
computation of $ind_{|\cF_X|}(\cF_X)$ will be simple.

We note that most of the $k$-dimensional subspaces have a Ferrers
diagram with a large number of dots. We will encode these
subspaces by the Ferrers tableaux form encoding and the other
subspaces by the extended representation encoding. We will decide
on a set $S_{\cF}$ of Ferrers diagrams which will be used for the
Ferrers tableaux form encoding. They will be taken by a decreasing
number of dots among all the Ferrers diagrams which can be
embedded in a $k \times (n-k)$ box.

We define  a new function $\widehat{Ind}$ in the following way:

\[\widehat{Ind}(X)=
\left\{ \begin{array}{cc}
 Ind_1(X)  & \cF_X \in S_{\cF}\\
Ind_2(X)+\Delta_X & \textrm{otherwise}\end{array},\right.\] where
$\Delta_X$ is the number of subspaces formed from $S_{\cF}$, which
are lexicographically succeeding $X$ by the extended
representation ordering. Similarly we will define an inverse
algorithm.

\section{Conclusion and Future Research}
\label{sec:conclude}

Three methods for enumerative encoding of the Grassmannian are
presented. The first is based on the Ferrers tableaux form of
subspaces. The second is based on the representation of subspaces
by their identifying vector and reduced row echelon form. The
complexity of the second method is superior on the complexity of
the first one. The third method which is a combination of the
first two reduces in average the constant in the first term of the
complexity for the second method. Improving on these methods is a
problem for future research.

Enumerative encoding of the Grassmannian is based on
representation and order of subspaces. Each such order defines a
lexicographic code~\cite{CoSl86} with prescribed minimum distance
(for two subspaces $X, ~Y \in \Gr$ the distance between $X$ and
$Y$ is defined by $d(X,Y) = \dim X + \dim Y - 2 \dim (X \cap
Y)$~\cite{KK}). It appears that some of these lexicodes are the
best known. For example, based of the Ferrers tableaux form
ordering we found a code with minimum distance 4 and size 4605 in
$\mathcal G_2(8,4)$ which is the largest known. Considering
lexicographic codes in the Grassmannian is a topic for future
research. There are some computational aspects involve in this
computation and this is currently under consideration.

\vspace{-0.1cm}

\section*{Acknowledgment}
This work was supported in part by the Israel Science Foundation
(ISF), Jerusalem, Israel, under Grant No. 230/08.


\end{document}